\documentclass[a4paper,11pt]{article}
\usepackage{jinstpub} 
\usepackage{lineno}
\usepackage{siunitx}
\usepackage{graphicx}
\usepackage{multirow}
\usepackage[T1]{fontenc}
\usepackage{url}
\usepackage{siunitx}

\title{\boldmath Development for the Belle II vertex detector upgrade with depleted monolithic active pixel sensors}

\author[o,1]{Y.~Onuki,\note{Corresponding author.}}

\author[a]{M.~Babeluk}
\author[a]{T.~Bergauer}
\author[a]{M.~Friedl}
\author[a]{C.~Irmler}
\author[a]{B.~Pilsl}
\author[a]{R.~Russo}
\author[a]{C.~Schwanda}

\affiliation[a]{Marietta Blau Institute for Particle Physics, Austrian Academy of Sciences, Dominikanerbastei 16, 1010 Vienna, Austria}

\author[b]{L.~Gaioni}
\author[b]{V.~Re}
\author[b]{E.~Riceputi}
\author[b]{G.~Traversi}

\affiliation[b]{Department of Engineering and Applied Sciences, University of Bergamo, Viale Marconi 5, I-24044 Dalmine (BG), Italy}

\author[c]{S.~Giroletti}
\author[c]{L.~Ratti}

\affiliation[c]{Department of Electrical, Computer and Biomedical Engineering, University of Pavia, Via Ferrata 5, I-27100 Pavia, Italy}

\author[d,e]{G.~F.~Benfratello}
\author[d,e]{S.~Bettarini}
\author[e]{F.~Bosi}
\author[d,e]{G.~Casarosa}
\author[e]{L.~Corona}
\author[d,e]{F.~Forti}
\author[e]{A.~Gabrielli}
\author[e]{M.~Massa}
\author[d,e]{L.~Massaccesi}
\author[e]{M.~Minuti}
\author[e]{A.~Moggi}
\author[d,e]{S.~Mondal}
\author[d,e]{G.~Rizzo}
\author[e]{M.~Rovini}
\author[e]{A.~Taffara}

\affiliation[d]{Dipartimento di Fisica “E. Fermi”, Universit\`a di Pisa, L.go B. Pontecorvo 3, I-56127 Pisa, Italy}
\affiliation[e]{INFN Sezione di Pisa, L.go B. Pontecorvo 3, I-56127 Pisa, Italy}

\author[f]{M.~Barbero}
\author[f]{P.~Barrillon}
\author[f]{R.~Boudagga}
\author[f]{P.~Breugnon}
\author[f]{D.~Fougeron}
\author[f]{P.~Pangaud}
\author[f]{J.~Serrano}
\author[f]{V.~Vobbilisetti}
\author[f]{D.~Xu}

\affiliation[f]{Aix Marseille Univ, CNRS/IN2P3, CPPM, Marseille,  France}

\author[g]{D.~Auguste}
\author[g]{J.~Bonis}
\author[g]{Y.~Peinaud}
\author[g]{M.~Winter}

\affiliation[g]{Laboratoire de Physique des 2 infinis Ir\`ene Joliot-Curie – IJCLab, Université Paris-Saclay, CNRS/IN2P3, IJCLab, 91405 Orsay, France}

\author[h]{J.~Baudot}
\author[h]{G.~Bertolone}
\author[h]{A.~Dorokhov}
\author[h]{G.~Dujany}
\author[h]{L.~Federici}
\author[h]{C.~Finck}
\author[h]{A.~Himmi}
\author[h]{C.~Hu-Guo}
\author[h]{A.~Kumar}
\author[h]{M.~Maushart}
\author[h]{F.~Morel}
\author[h]{H.~Pham}
\author[h]{I.~Ripp-Baudot}
\author[h]{R.~Sefri}
\author[h]{P.~Stavroulakis}
\author[h]{I.~Valin}

\affiliation[h]{Universit\'e de Strasbourg, CNRS, IPHC UMR 7178, F-67000 Strasbourg, France}

\author[i]{F.~Bernlochner}
\author[i]{C.~Bespin}
\author[i]{J.~Dingfelder}
\author[i]{T.~Kishishita}
\author[i]{H.~Krüger}
\author[i]{L.~Schall}
\author[i]{M.~Vogt}

\affiliation[i]{Physikalisches Institut, Rheinische Friedrich-Wilhelms-Universität Universität Bonn, Nussallee 12, 53115 Bonn, Germany}

\author[j]{M.~Karagounis}

\affiliation[j]{University of Applied Sciences and Arts Dortmund, Sonnenstraße 96-100, 44139 Dortmund, Germany}

\author[k]{Y.~Buch}
\author[k]{A.~Frey}
\author[k]{B.~Schwenker}
\author[k]{M.~Schwickardi}

\affiliation[k]{II. Physikalisches Institut, Georg-August-Universität Göttingen, Friedrich-Hund-Platz 1, 37077 Göttingen, Germany}

\author[l,m]{K.~Hara}
\author[l,m]{D.~Jeans}
\author[l,m]{K.~R.~Nakamura}
\author[l,m]{Y.~Okazaki}

\affiliation[l]{High Energy Accelerator Research Organization (KEK), Tsukuba 305-0801, Japan}
\affiliation[m]{The Graduate University for Advanced Studies (SOKENDAI), Hayama 240-0193, Japan}

\author[n]{T.~Higuchi}

\affiliation[n]{Kavli Institute for the Physics and Mathematics of the Universe (WPI), University of Tokyo, Kashiwa-no-ha 5-1-5, Kashiwa 277-8583, Japan}

\author[o]{S.~Wang}

\affiliation[o]{Department of Physics, University of Tokyo, Hongo 7-3-1, Tokyo 113-0033, Japan}

\author[p]{C.~Lacasta}
\author[p]{C.~Marinas}
\author[p]{J.~Mazorra de Cos}
\author[p]{L.~Molina-Bueno}

\affiliation[p]{Instituto de Fisica Corpuscular (IFIC), CSIC-UV, Catedratico Jose Beltran, 2. E-46980 Paterna, Spain}

\author[q]{A.~Bevan}
\author[q]{M.~Bona}
\author[q]{D.~Howgill}

\affiliation[q]{School of Physical and Chemical Sciences, Department of Physics and Astronomy, Queen Mary University of London, 327 Mile End Road, London, E1 4NS, United Kingdom}

\author[r]{W.~ Song}
\author[r]{J.~Gong}
\author[r]{X.~Gao}

\affiliation[r]{College of Physics, Jilin University, 2699 Qianjin Street, Changchun, Jilin, China}

\author[s]{A.~Fernandez~Prieto}
\author[s]{A.~Gallas~Torreira}

\affiliation[s]{Universidade de Santiago de Compostela, 2010 Instituto Galego de Física de Altas Enerxías (IGFAE), Colexio de San Xerome, PZ Obradoiro, S/N. E-15782 Santiago de Compostela, Spain }

\emailAdd{yoshiyuki-onuki@g.ecc.u-tokyo.ac.jp}

\abstract{
The vertex detector upgrade project for the Belle II experiment, based on CMOS depleted monolithic active pixel sensor technology, is planned to be carried out in conjunction with the major modification of the interaction region of the SuperKEKB collider during Long Shutdown 2 (2032–2034). 
The MAPS sensor, named OBELIX currently under development, is derived from the successor to TJ-Monopix2, with modifications implemented to ensure compatibility with the Belle II trigger system. 
The new vertex detector consists of two layers of four self-supported consecutive OBELIX sensors, and three layers of discrete OBELIX sensors mounted on mechanical support structures with readout flex circuits attached to the sensors. The detector is arranged cylindrically around the beam pipe at radii ranging from 14 mm to 140 mm.
The minimization of the material budget is required in order to enhance physics performance. We present an overview of the project and its latest developments, with particular emphasis on 
the development of low-material-budget flex circuits employing aluminum conductors.
}


\begin{document}
\begin{NoHyper}
\maketitle
\end{NoHyper}
\flushbottom

\section{Introduction}
\label{sec:intro}

The Belle II experiment \cite{Abe:2010gxa} at KEK in Tsukuba, Japan provides rich flavor physics programs of beauty, charm, light quarks and tau-lepton to discover phenomena beyond the standard model. 
It uses the asymmetric energy e$^+$e$^-$ collider SuperKEKB \cite{Ohnishi2013}, which has been taking data at the energy of $\Upsilon(4S)$ resonance and nearby energies since 2019. 
SuperKEKB achieved the world's highest peak luminosity $5.24 \times 10^{34}$ $\rm{cm^{-2}}\rm{s}^{-1}$, and delivered a total of \SI{750}{fb^{-1}} by April 2026.
SuperKEKB is the world’s first accelerator to employ the nano-beam scheme to achieve the target peak luminosity $6\times10^{35}$ $\rm{cm^{-2}}\rm{s^{-1}}$ and the target integrated luminosity \SI{50}{ab^{-1}}.
The current Belle II vertex detector (VXD) is composed of two layers of DEPFET-type \cite{pxd} pixel detector (PXD) and four layers of double-sided silicon strip detector (SVD) with chip-on-sensor and conventional hybrid readout schemes \cite{Belle-IISVD:2022upf}.





\section{VTX upgrade project}
\label{sec:ls2}

\begin{figure}[htbp]
    \centering
    \includegraphics[trim= 10 8 10 35,clip, width=0.65\linewidth]{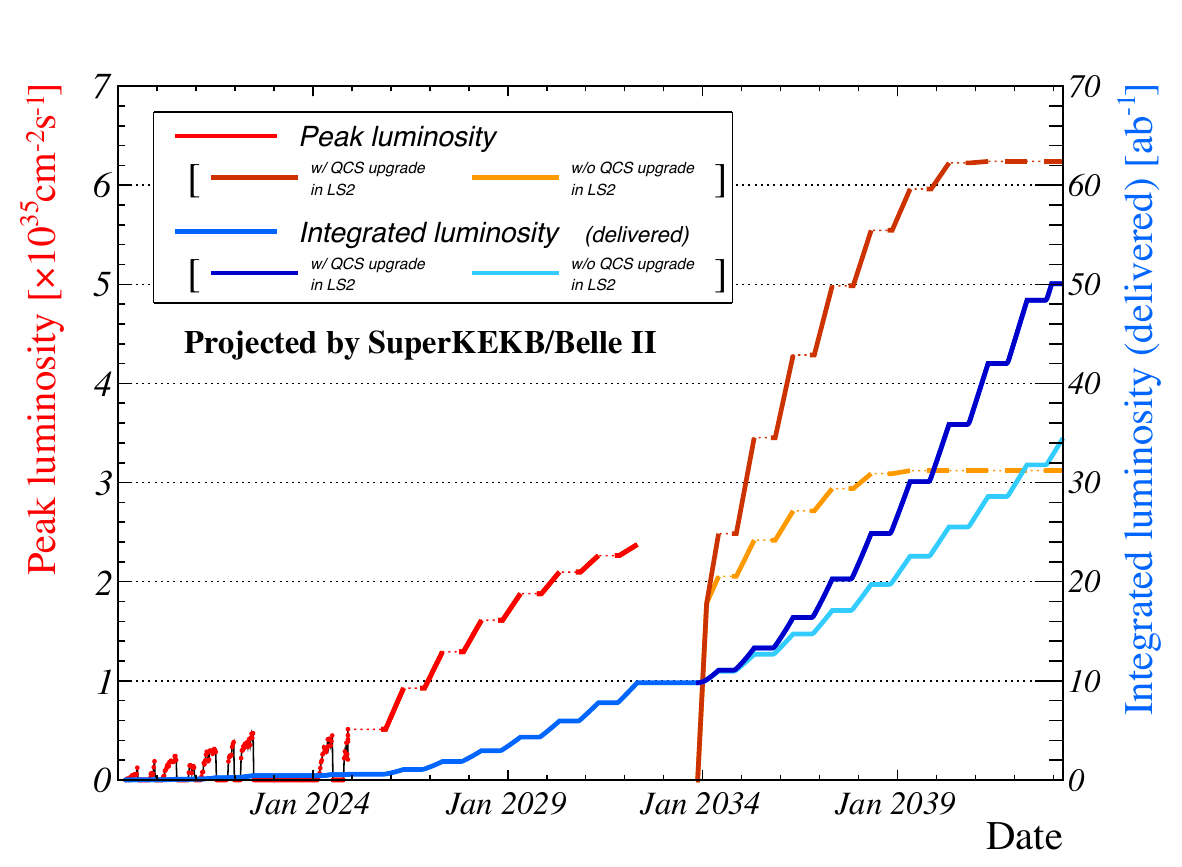}
    \caption{Luminosity projection plot of SuperKEKB and Belle II experiment with and without QCS upgrade.}
    \label{fig:LuminosityProjection}
\end{figure}

\begin{figure}
    \centering
    \includegraphics[trim= 50 170 50 170,clip, width=0.75\linewidth]{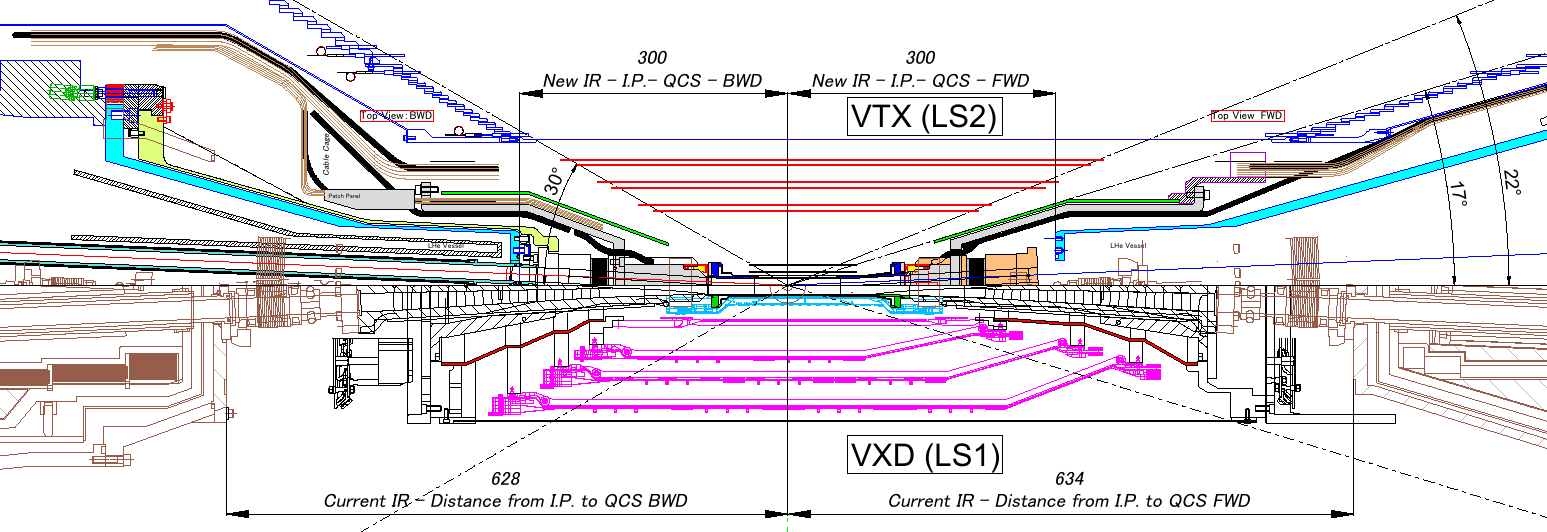}
    \caption{Schematic layout of two IR regions before and after the LS2 in comparison. Lower half shows the current IR region with VXD since LS1. Upper half shows the upgraded IR region with VTX after LS2.}
    \label{fig:IRComparison}
\end{figure}

\begin{figure}[htbp]
    \centering
    \scalebox{0.95}{
    \includegraphics[trim= 10 10 10 10,clip,width=.55\textwidth]{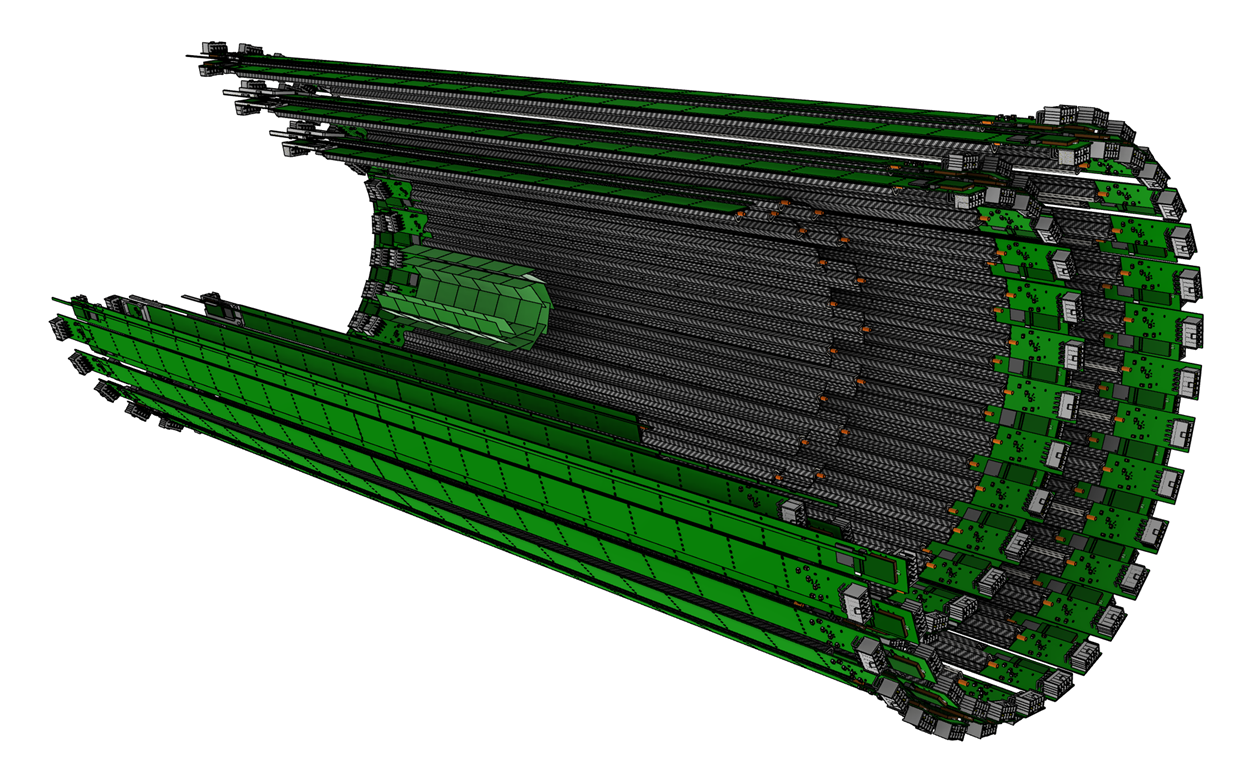}
    }
    \qquad
    \includegraphics[trim= 40 10 50 10,clip, width=.33\textwidth]{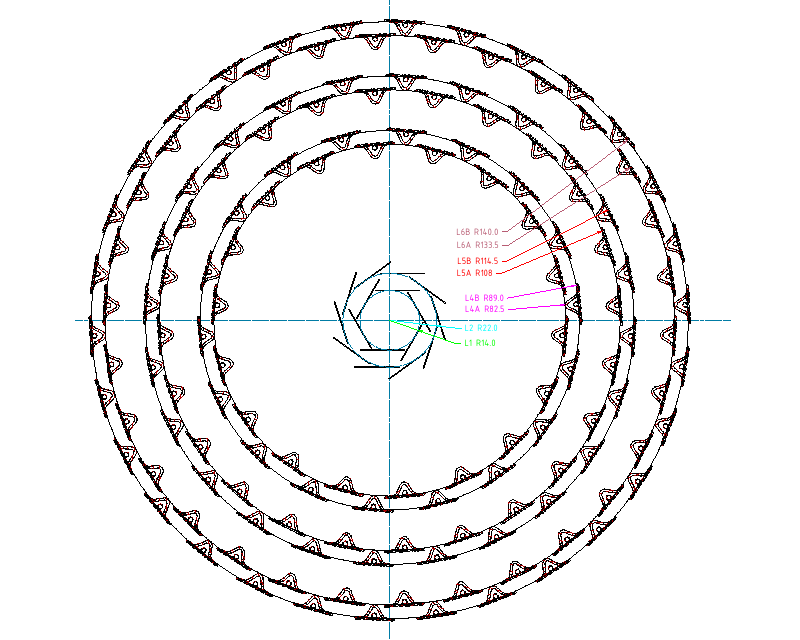}
    \caption{Schematic view and cross sectional view of VTX.}
    \label{fig:VTXviews}
\end{figure}

The present SuperKEKB is expected to reach a peak luminosity of $2\times10^{35}$ $\rm{cm^{-2}} \rm{s^{-1}}$. 
However, in order to achieve the target peak luminosity and the integrated luminosity, an upgrade of the current beam interaction region (IR) including final focusing superconducting quadrupole magnets (QCS) will be required. 
Figure \ref{fig:LuminosityProjection} shows the luminosity projection plots of SuperKEKB and the Belle II. 
In this plot, 
one can see two long shutdowns: the first one in 6/2022-12/2024 (LS1), during which various improvements were performed on the accelerator and detector including the completion of PXD, and a second one, planned in 2032-2034 (LS2), during which the IR and the detector will be upgraded.
The enlarged space envelope of the upgraded IR and QCS is incompatible with the VXD geometry since the new QCS magnets,  
together with the extended compensation magnets for the 1.5 T magnetic field induced by the Belle II superconducting solenoid magnet, are located $\sim$ 100 mm closer to the collision point.
To cope with the space-limited IR and the harsh beam background conditions expected after the upgrade, a fully pixelated vertex detector (VTX) with depleted monolithic active pixel sensors, 
is proposed as one of the major upgrades \cite{Aihara:2024cdr} of Belle II.
Figure~\ref{fig:IRComparison} shows a schematic drawing of the present IR after LS1 in the lower half and foreseen upgrade after LS2 in the upper half. 
Conical shapes at both ends of VTX/VXD show the envelopes of QCS cryostat. 
The VTX comprises five cylindrical layers of OBELIX sensors arranged around the beam pipe at radii from \SI{14}{mm} to \SI{140}{mm}, providing the polar-angle acceptance from 22 $^\circ$ to 150 $^\circ$. 
Each layer is uniformly tiled with OBELIX chips of the same design and size, arranged in ladders, whose shape corresponds to the staves of a barrel.
The configuration of each layer is summarized in Table~\ref{tab:vtxlayer} and the schematic layout is shown in Figure~\ref{fig:VTXviews}. 

\begin{table}
    \centering
    \caption{VTX configuration in terms of layer, radius, number of ladders, modules and sensors, where a/b in L3-5 shows staggered layout in each layer.}
    \scalebox{0.9}{
    \begin{tabular}{c|c|c|c|c|c}
        Layer            & L1   & L2   & L3a/b        & L4a/b          & L5a/b\\
        \hline
        Configuration    & \multicolumn{2}{c|}{iVTX}    & \multicolumn{3}{c}{oVTX}\\
        \cline{2-3}
        \cline{4-6}
        Radius           & 14.0 & 22.0 & 82.5/89.0 & 108.0/114.5 & 133.5/140.0\\
        \# Ladders       &  6   & 10   & 36        & 48          & 60\\
        \# Sensor/module &  4   & 4    & 12/13     & 16/17       & 20\\
    \end{tabular}}
    \label{tab:vtxlayer}
\end{table}

The ladder of the inner two layers in L1-2 (iVTX) is composed of four consecutive self-supported OBELIX sensors interconnected with a post-process redistributed layer (RDL). 
The iVTX ladders are mounted on the beam pipe and organized in a windmill structure. 
The ladders of the outer three layers, L3-5 (oVTX), consist of OBELIX chips mounted on omega-shaped support made of lightweight carbon-fiber, including a cold plate and liquid-cooling pipes. 
The OBELIX chips of a ladder are wire bonded to a flex PCB mounted above the chips, which provides electrical connections.
The oVTX ladders are mounted on the end rings, which are part of the mechanical support structure of the VTX, and are arranged in a staggered layout.

\section{OBELIX sensor}
\label{sec:obelix}
\begin{figure}
    \centering
    \scalebox{0.75}{
        \begin{tabular}[b]{|c|c|}\hline
            Pitch                      & 33 $\mu \rm{m}$ \\ 
            \hline
            Signal ToT                 & 7 bits \\
            \hline
            Time stamping              & 50 to 100 $ \rm{ns}$\\
            \hline
            \begin{tabular}{c}
                Fine time *\\
                stamping\\
            \end{tabular}
            &    
            \begin{tabular}{c}
                $\sim$ 5 \rm{ns} \\
                for hit rate < 10 MHz/c$\rm{m^2}$\\
            \end{tabular}\\
            \hline
            Trigger handling           & 30 kHz with \rm{10} $\mu \rm{s}$ delay\\
            \hline
            \begin{tabular}{c}
                Trigger *\\
                output\\
            \end{tabular}
            & 
            \begin{tabular}{c}
                $\sim$ 10 \rm{ns} resolution\\
                with low granularity\\
            \end{tabular}\\
            \hline
            \begin{tabular}{c}
                Power\\
                (with hit rate)\\
            \end{tabular}
            & 
            \begin{tabular}{c}
                200 to 300 \rm{mW/$cm^{2}$}\\
                (1 to 120 \rm{MHz/$cm^{2}$}\\
            \end{tabular}\\
            \hline
            Bandwidth & 1 output 320 MHz\\
            \hline
            \multicolumn{2}{r}{* optinal features}\\
        \end{tabular}
    }
    \qquad
    \scalebox{0.95}{
    \includegraphics[trim= 3 3 3 3,clip, width=0.5\linewidth]{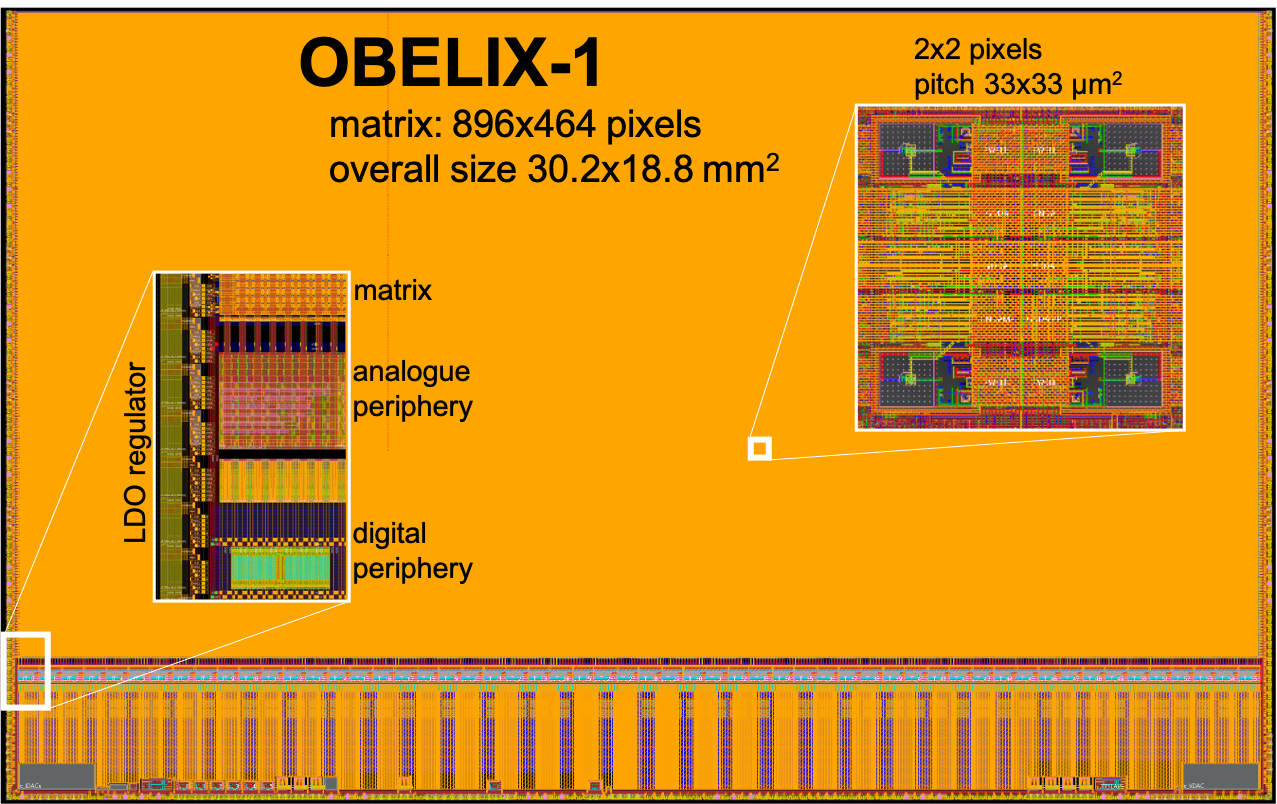}
    }
    \caption{OBELIX-1 chip floor plan. Left table shows the specification based on the simulation.}
    \label{fig:obelix}
\end{figure}

The OBELIX sensor \cite{babeluk_cmos_2024}, derived from TJ-Monopix2 \cite{moustakas_design_2021} originally developed for the outer layers of the ITk \cite{itk} in ATLAS experiment, is based on depleted monolithic active pixel sensor in 180 nm CMOS technology and features a matrix of $896 \times 464$ pixels with $\SI{33}{\mu m}$ pitch.
The OBELIX chip can withstand a non-ionizing energy loss (NIEL) fluence $5 \times 10^{14}$ 1 MeV $\rm{n_{eq}}/\rm{cm^{2}}$ and a total ionizing dose of \SI{100}{Mrad}.
The digital periphery needed to be largely redesigned and features the trigger unit with a dual-stage memory buffer to allow operation with the external Belle II trigger.
It is able to handle hit rates up to 120 MHz$/\rm{cm^2}$ with a trigger latency of 10 $\mu \rm{s}$ and \SI{30}{kHz} trigger rate. 
Figure~\ref{fig:obelix} shows the floor plan of OBELIX with the features summarized in the left table.
The analogue signal caused by incident particle is processed by the DC-coupled cascode front-end circuit, which was selected among four front-end circuit designs tested with TJ-Monopix2\cite{alice, Yanik}. 
The chip provides also a 7-bit Time-Over -Threshold, with a time stamping of 50-100 ns.
Moreover, the threshold can be tuned on a pixel-by-pixel basis with a 3-bit tuning circuit to compensate threshold variations across the pixel matrix.
The OBELIX features a serial output to the Belle II trigger system for the transmission of low-latency, coarse-granularity macropixel hit data based on pixel HitOr signals, referred to as the Track Trigger Transmission (TTT) module, and a precision timing module called the Periphery Time-to-Digital (PTD) which provides precision timing information. 
Both are foreseen to be enabled only in oVTX but not in iVTX, due to the higher occupancy and power consumption constraints \cite{babeluk_cmos_2024}.
On-chip low-dropout regulators (LDOs) are implemented in the periphery to allow operation with an input supply voltage range of 2-3 V to compensate for the expected voltage drops due to the long supply rails along the flex.

\section{iVTX}
\label{sec:iVTX}
The iVTX layers are designed to satisfy the target material budget of 0.2 \% $\rm{X}/\rm{X_0}$ per layer requested by physics performance study \cite{Aihara:2024cdr}.  
For each iVTX ladder, a block of four consecutive OBELIX sensors will be diced from the same wafer. 
The RDL will be integrated on the top surface to provide interconnections between the chips and a flex glued at the edge to send the data to backend \cite{max}. 
The assembly will then undergo a backside thinning process down to a thickness of 50–100 $\mu\mathrm{m}$, as shown in Figure 5, or slightly thicker to ensure sufficient mechanical stiffness.
The cooling of the "all-silicon ladder" is foreseen as a passive cooling using a thin layer of Thermal Pyrolytic Graphite (TPG) having 1500 \rm{W/m$\cdot$K} conductance, already used in CBM experiment \cite{cbmvtx}, glued to the sensor. 
The TPG, acting as heatsink and mechanical structure to secure self-supported stiffness, is thermally connected to end-mount blocks, which are actively cooled by a coolant flowing through micro-channels inside. 
The validation of the passive cooling scheme using a mockup iVTX ladder with dummy silicon and actual TPG is under preparation.


\begin{figure}
    \centering    
    \includegraphics[width=0.75\linewidth]{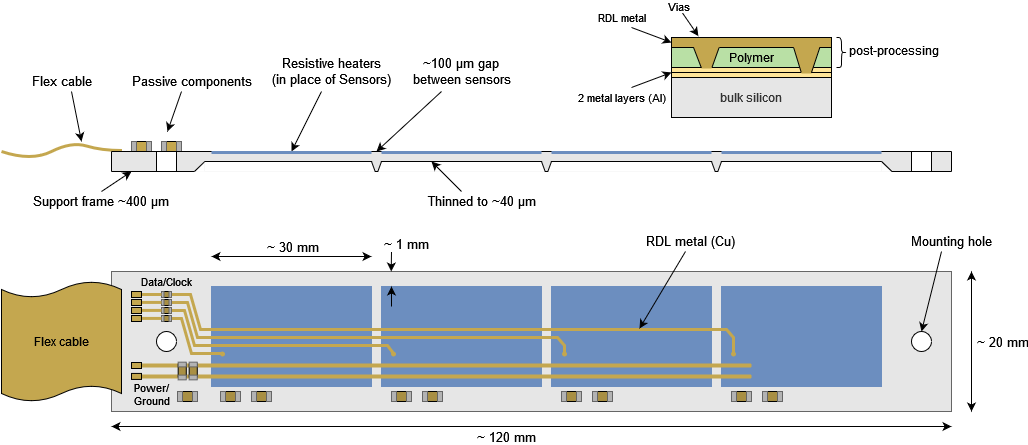}
    \caption{iVTX ladder configuration.}
    \label{fig:ivtx}
\end{figure}

\section{oVTX}
\label{sec:oVTX}
\begin{figure}
    \centering
    \includegraphics[trim= 5 5 5 5,clip,width=0.55\linewidth]{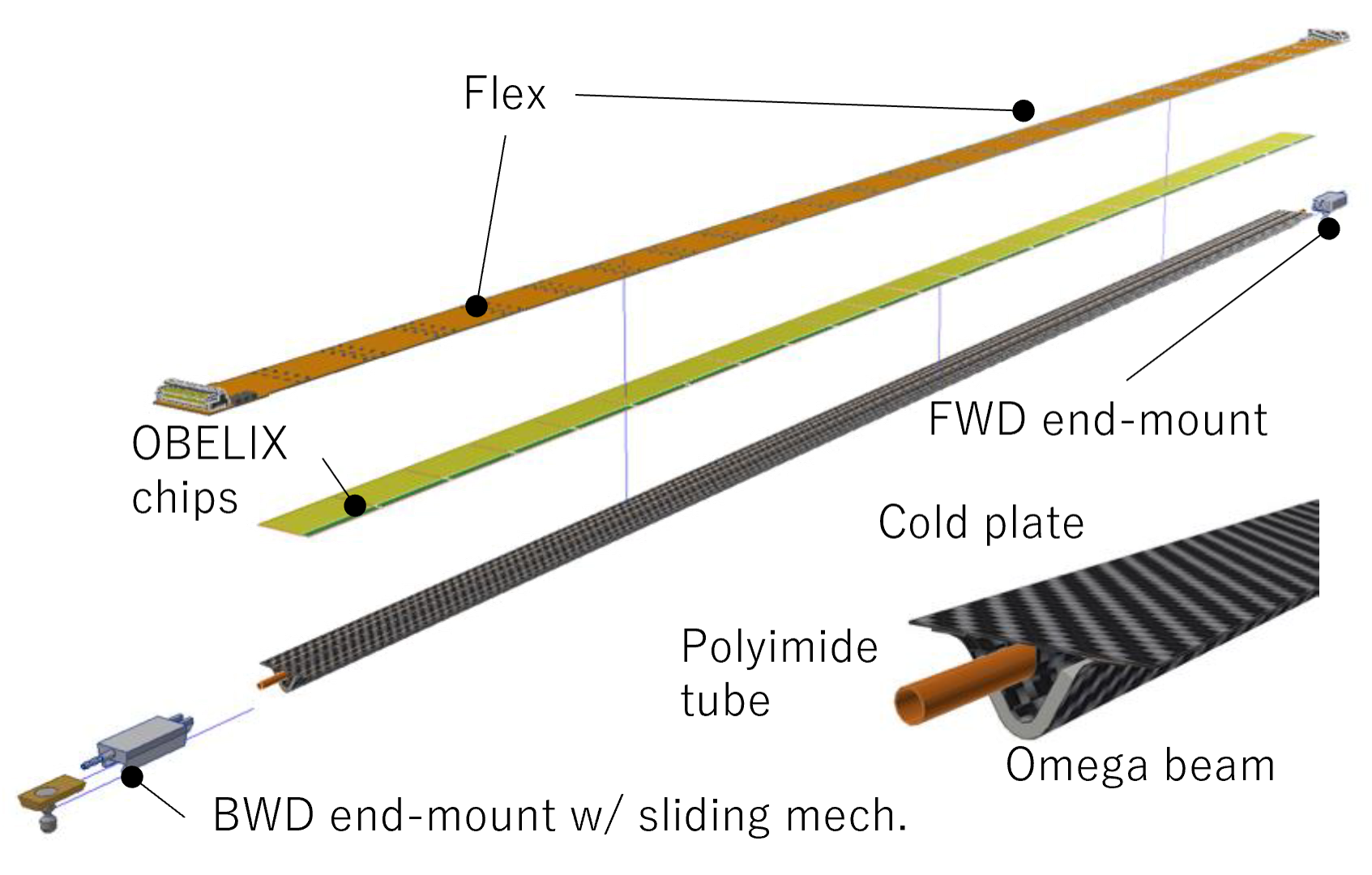}
    \caption{Schemativ view of oVTX ladder configuration.}
    \label{fig:ovtx}
\end{figure}

\begin{figure}
    \centering
    \includegraphics[trim= 5 6 5 6, clip, width=0.75\linewidth]{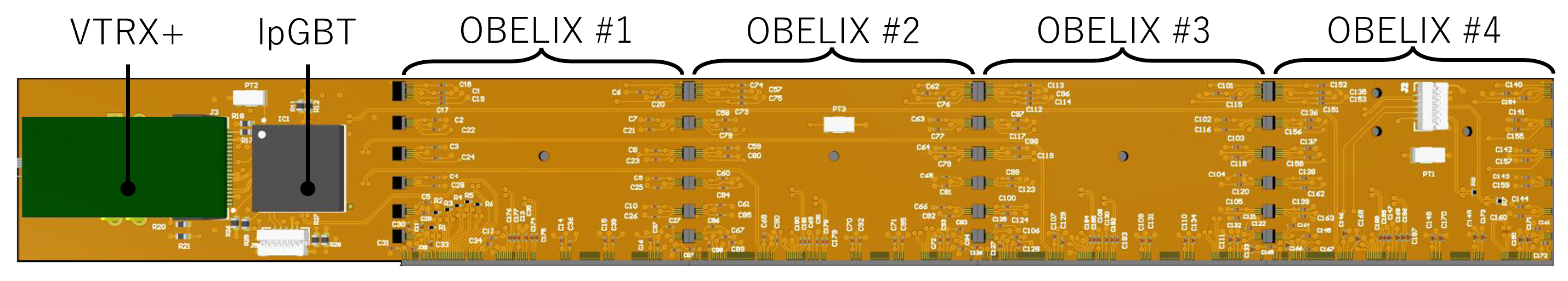}
    \caption{Schematic view of prototype copper flex to verify the scheme of four OBELIX sensors to readout through the lpGBT and VTRX+.}
    \label{fig:protoflex}
\end{figure}

The oVTX ladder is designed to satisfy target material budget of 0.8 \% $\rm{X}/\rm{X_0}$ per layer requested by physics performance study \cite{Aihara:2024cdr}. 
The design of oVTX ladders, strongly inspired by the ALICE ITS2 \cite{fantoni_upgrade_2020}, consists of thinned OBELIX sensors, a support structure with end-mounts in both ends, two readout flex
\cite{ovtx} shown in Figure~\ref{fig:ovtx}. 
The OBELIX sensors in the oVTX will be thinned to 50–100 $\mu\mathrm{m}$, presumably thinner than those in the iVTX, because the support structure is prepared separately.
The support structure comprises an omega beam, a cold plate, and forward and backward end-mounts. 
The omega beam consists of multilayer carbon fibers combining unidirectional and plain-weave fabrics, 
while the cold plate is made of a high-thermal-conductivity carbon substrate with an embedded polyimide tube. 
The end-mounts are made of stainless steel and serve as interface between the VTX structure and the ladder, with the forward end-mount incorporating a mechanism to compensate for thermal expansion and shrinkage of the ladder.
One or two flex per ladder supply the power voltage and transmit control and data signals to each OBELIX chip, as described in the next section.
The most challenging elements of the oVTX in terms of mechanics and powering are the layer 5 ladders, being both the longest and the most numerous.
Modularization and mass production compatible with the global schedule are currently under investigation.
Another particularly challenging aspect of the oVTX is the very limited service space allocated to the ladders. 
The connecting cable between the flex and the backend electronics must have a small cross section to feed through the available space and should ideally aggregate multiple lines.
The readout will be performed using rad-hard components developed for HL-LHC experiments: the Low Power GigaBit Transceiver (lpGBT) \cite{Moreira2025lpGBTLR} and the Versatile Transceiver plus (VTRx+) \cite{vtrx} optical link. 
A prototype flex using lpGBT and VTRx+ to read four OBELIX chips is being developed to cope with the small available service space as shown in Figure~\ref{fig:protoflex}.  

\section{Aluminum conductor flexible circuit board}
\label{sec:alflex}
The tracking performance requires a low material budget per layer, which not only depends on the sensor thickness but also mainly on the flex, providing power and readout signal by copper conductors.
There is a pioneering work of the low material budget flex for ATLAS IBL using Aluminum conductor developed by CERN PCB workshop \cite{ibl}. 
They fabricate the flex composed of four layers of Cu conductor and two layers of Al conductor connected by power vertical interconnect accesses (VIAs) processed by Physical Vapor Deposition (dry plating) of 0.2 $\mu$m Cr and 2 $\mu$m Cu followed 25 $\mu$m Cu electroplating (wet plating) to connect the Al layer with Cu layer. 
This technology could enable flex circuits with routing conductors made predominantly of aluminum, while limiting the use of copper to localized VIAs, thereby reducing the material budget by up to 0.6 \% X/$\rm{X_{0}}$ per layer in oVTX.
We are developing an Al flex in collaboration with the CERN using the same design as shown in the Figure~\ref{fig:protoflex} and expecting the CERN as a main supplier of the Al flex.

\begin{figure}
    \centering
    \includegraphics[trim= 5 10 5 22,clip,width=0.75\linewidth]{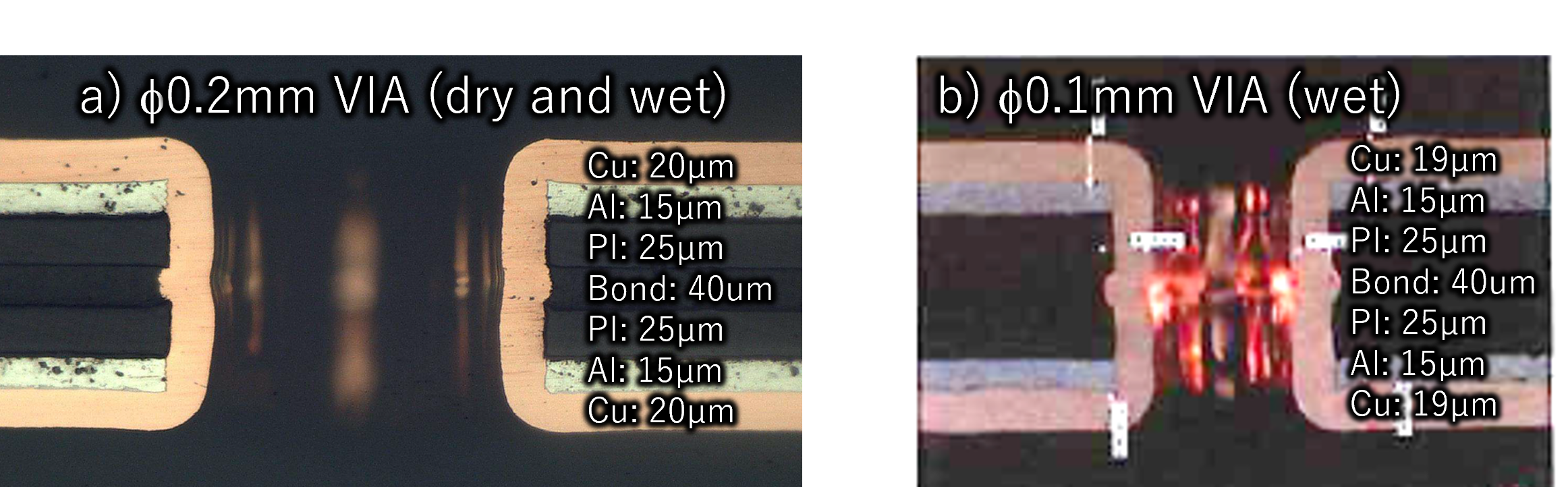}
    \caption{Cross sections of VIA formed in stacked AlPI. a) are fabricated by similar way of CERN in Japan. b) is a new way of wet only process. Each thickness of the components are also shown.}
    \label{fig:via}
\end{figure}

\begin{figure}
    \centering
    \includegraphics[trim= 5 8 5 5,clip,width=0.8\linewidth]{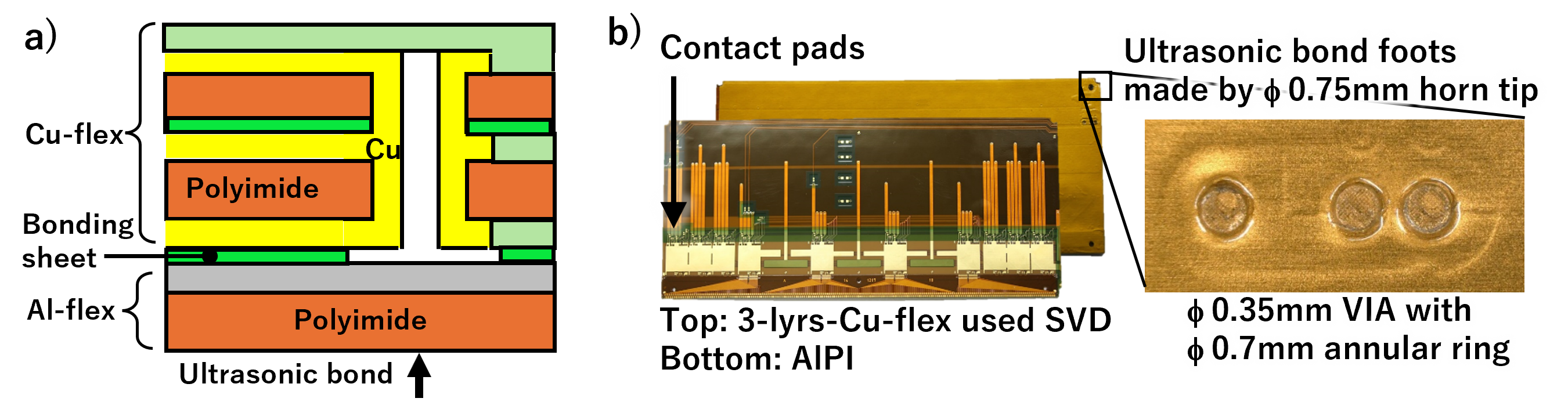}
    \caption{a) shows a schematic view of ultrasonic bonding applied to the flex shown in b). (b) is an image of the Cu–Al flex and the magnified bonding foot.}
    \label{fig:ultrasonic}
\end{figure}

We are also exploring a similar approach in collaboration with a Japanese supplier using both dry and wet plating techniques, as well as two alternative new approaches: a wet-only plating process without dry plating, and an ultrasonic bonding. Both approaches currently use unetched Al-Polyimide (AlPI) to focus on VIA formation.
Figure~\ref{fig:via} a) shows the cross sections of 0.2 mm diameter VIAs of two stacked AlPI films, fabricated using a method similar to that used at CERN, in which 0.2 $\mu$m Cr and 2 $\mu$m Cu are deposited inside the VIA by ion plating, followed by Cu electroplating.
Figure~\ref{fig:via} b) fabricated by wet-only plating using a double zincate process, followed by electroless plating of 1 $\mu$m Ni and 4 $\mu$m Cu, and then electroplating Cu up to a total thickness of $\sim$ 20 $\mu$m.
In terms of production cost, ion plating accounts for a large portion of the total cost in Japan, whereas the new wet-only process is expected to offer a lower production cost.
Another approach to achieve a low material budget flex is to use copper for the signal layers with fine-pitch traces, while using aluminum for the power layers with large-area conductors. 
This Cu-Al flex has a material budget roughly 10 \% higher than that of the all-Al flex while still providing a significant reduction in material budget.
To explore the feasibility of a process without dry plating from a cost-reduction perspective, we tested bonding between Cu VIA, with the same Cu flex used in current SVD, and AlPI using a ultrasonic bonding technique in Japan.
Figure~\ref{fig:ultrasonic} a) shows the cross section of Al-Cu flex, which combines Cu and Al flexes using a bonding sheet followed by ultrasonic bonding through the polyimide. 
Figure~\ref{fig:ultrasonic} b) shows images of actual Cu-Al flex, combining the Cu-flex used in current SVD and AlPI, with the inset showing the VIA location and the ultrasonic bonding region.
A 0.8 mm diameter VIA fabricated using the wet-only process and a 0.35 mm diameter VIA fabricated by ultrasonic bonding both demonstrated electrical conduction, with preliminary resistance values of the order of $\mathrm{m}\Omega$ per VIA. These values are considered sufficiently low for practical circuit applications.

\section{Conclusions}

The SuperKEKB accelerator is planning a major upgrade to reach its target luminosity of 
$6\times10^{35}$ $\rm{cm^{-2}}\rm{s^{-1}}$ during the long shutdown 2 in 2032.
The Belle II vertex detector will also be upgraded to improve the vertex performance and robustness against high beam backgrounds,  and cope with the new geometry of the interaction region.
The VTX will be using the OBELIX DMAPs chip, whose first prototypes has recently been sent to production. 
The aluminium conductor flex is one of the key components that has to be optimized to reduce the material budget in order to improve low momentum particle tracking. Several approaches are being explored, in collaboration with CERN and Japanese suppliers.


\acknowledgments
This work has received the support from the European Union’s Horizon 2020 Research and Innovation programme under Grant Agreements no 101004761 (AIDAinnova),  Multilateral Scientific and Technological Cooperation in the Danube Region (MULT 03/23),
Horizon 2020 ERC-Consolidator Grant No. 819127,
TY-FJPPN (Toshiko Yuasa-France Japan Particle Physics Network), 
the MCIU with funding from the European Union NextGenerationEU (PRTR-C17.I01),
Generalitat Valenciana (GVANEXT), Project ASFAE/2022/016,
JSR–UTokyo Collaboration Hub CURIE, and The Precise Measurement Technology Promotion Foundation (PMTP-F).






\bibliographystyle{elsarticle-num}
\bibliography{biblio}




\end{document}